\documentstyle[12pt,a4wide,epsf]{article}

\begin{document}
\renewcommand{\thefootnote}{\fnsymbol{footnote}}
\makebox[2cm]{}\\[-1in]
\begin{flushright}
\begin{tabular}{l}
TUM--T31--74/94\\
hep--ph/9409247
\end{tabular}
\end{flushright}
\vskip0.4cm
\begin{center}
{\Large\bf
Theory of Semi-- and Nonleptonic Decays of\\[4pt]
Heavy Mesons\footnotemark}
\footnotetext{Invited talk given at ICHEP 94, Glasgow, Scotland,
July 1994; to appear in the Proceedings.}

\bigskip

Patricia Ball

\bigskip

{\em Physik--Department, TU M\"{u}nchen, D--85747 Garching, Germany}

\bigskip

{\em September 7, 1994}

\bigskip

{\bf Abstract:\\[5pt]}
\parbox[t]{\textwidth}{\small
I review some of the recent developments in the theoretical
description of weak inclusive decays of heavy mesons. The topics cover
the value of $|V_{cb}|$ as extracted from semileptonic inclusive
decays and a short discussion of the theoretical errors. I also present
the results of a recent calculation of next--to--leading order
corrections to nonleptonic inclusive B decays which allows an
improved prediction of the semileptonic branching ratio of B mesons.
}
\end{center}

\smallskip

\section{Outline of Theoretical Foundations}

During the recent two years, the theoretical description of inclusive
decays of heavy hadrons has experienced considerable progress. For
quark masses $m_Q\gg \Lambda_{\mbox{\scriptsize QCD}}$, the
well--known short--distance expansion technique yields an expansion in
inverse powers of the heavy quark mass, the so--called heavy quark
expansion (HQE) \cite{general}. The starting point for the HQE of,
e.g., the decay rate of a B meson into a final state X is its
representation as
imaginary part of the relevant forward--scattering amplitude:
\begin{equation}\label{eq:1}
\Gamma(B\to X) = \frac{1}{m_B}\,{\rm Im}\,i\!\!\int\!\! d^4\! x\,
\langle B|T{\cal L}_W(x) {\cal L}_W(0)|B\rangle.
\end{equation}
Here ${\cal L}_W$ is the effective weak Lagrangian mediating the decay
$B\to X$. As shown
in \cite{general}, for a very heavy b quark mass $m_b$, a short
distance expansion of Eq.\ (\ref{eq:1}) yields:
\begin{eqnarray}
\lefteqn{\Gamma(B\to X) = \frac{G_F^2m_b^5}{192\pi^3}\,|
{\rm CKM}|^2\left\{
C_0^{(X)}(\alpha_s(\mu)) \,\frac{\langle B|\bar b b|B\rangle}{2m_B}
\right.} \nonumber\\
& & \kern-0.6cm \left. {} + C_2^{(X)}(\alpha_s(\mu))\,\frac{1}{m_b^2}\,
\frac{\langle B|\bar b g_s \sigma_{\mu\nu} F^{\mu\nu}
b|B\rangle}{2m_B} + {\cal O}\left(\!\frac{1}{m_b^3}\!\right)\right\}.
\label{eq:2}
\end{eqnarray}
Here CKM denotes the appropriate CKM matrix elements, the $C_i^{(X)}$
are short--distance Wilson--coefficients, which depend on the parton
model process underlying the decay $B\to X$. $b$ is the
b quark field in full QCD, $m_B$ is the mass of the B meson
and $F^{\mu\nu}$ the gluonic field--strength tensor.

Without going into too much details, let me mention just a few
general features of the above expansion.
First we remark that HQE strongly resembles deep inelastic
scattering with the difference
that the value of the expansion parameter $m_b$ is fixed and cannot be
controlled by the experimenter, so that higher twist effects in $1/m_b$
are important. On the other hand, the corresponding hadronic matrix
elements can be expressed as moments of a universal distribution
function \cite{distri,SVlimit} and are thus measurable, at least in
principle, cf.\ \cite{grozin}.

The first term in the above series, $\langle B|\bar b b|B\rangle$,
just reproduces the free quark decay process; non--perturbative
corrections to that picture are suppressed by terms of ${\cal
O}(1/m_b^2)$ or higher. There are no terms of order $1/m_b$, since
all possible gauge--invariant operators of suitable dimension either
vanish or can be reduced to $\bar b b$ by the equations of motion.

The Wilson coefficients $C_j^{(X)}(\alpha_s(\mu))$ encode the short
distance behaviour of Eq.\ (\ref{eq:1}) and are calculable
within perturbation theory; they depend on the parton content of X, the
renormalization scale $\mu$ and the renormalization scheme. In
particular, $C_0^{(X)}$ contains the radiative corrections to the
free quark decay and has been studied for various processes
\cite{CM78,ACMP81,HP,nir,tum67}.

Let me also mention some words of caution. In order to separate
clearly the expansion in $1/m_Q$ from the one in $\alpha_s$, the HQE
has to be done in terms of the {\em pole mass}. This mass definition,
however, is unphysical and contradicts the confinement property of
QCD. It has been shown,
that this contradiction reflects itself in an intrinsic
ambiguity of the definition of the pole mass which is said to be caused
by a ``renormalon'' \cite{ren1}. Although it was shown in
\cite{ren2} that the renormalon cancels in the semileptonic rate, at
least up to order $1/m_b^2$, there is still a number of questions to
be answered as far as nonleptonic and exclusive decays are concerned.

\section{Applications I: $|$V$_{cb}|$ from
Semileptonic Inclusive Decays}

One immediate application of the HQE is the extraction of $|V_{cb}|$
from semileptonic inclusive decays. The decay rate $\Gamma(B\to X_c
e\nu)$, of the generic form of Eq.\ (\ref{eq:2}), contains to order
$1/m_b^2$ five unknown parameters: $V_{cb}$, $m_b$, $m_c$ and two
hadronic matrix elements:
\begin{eqnarray}
2m_B\lambda_1 & = & \langle\,B\,|\,\bar{b}_v (iD)^2
b_v\,|\,B\,\rangle, \nonumber\\
6m_B\lambda_2 & = & \langle\,B\,|\,\bar{b}_v \frac{g}{2}\,
\sigma_{\mu\nu} F^{\mu\nu}b_v\,|\,B\,\rangle,
\end{eqnarray}
where $b_v$ is defined as $b_v = e^{im_b v x}b$ and $v_\mu$ is the
four--velocity of the B meson.

Whereas $\lambda_2$ is directly related to the observable
spectrum of beautiful mesons,
\begin{equation}
\lambda_2 \approx \frac{1}{4}\,(m_{B^*}^2 - m_B^2)= 0.12\,{\rm GeV}^2,
\end{equation}
the quantity $\lambda_1$ is difficult to measure, cf.\
\cite{grozin}. Physically, $-\lambda_1/(2m_b)$ is
just the average kinetic energy of the b quark inside the meson.
At present, only a QCD sum rule estimate is available,
according to which $\lambda_1 \simeq -0.6$ GeV$^2$ \cite{BB94}.
 This result has been met with caution
(see, e.g.\ \cite{vir}), since it corresponds in fact to a surprisingly
large momentum of the b quark inside the meson of order
(700--800)$\,$MeV.
However, in a recent series of papers, cf.\ \cite{grozin,SVlimit},
an upper bound on $\lambda_1$ was derived, to wit
$\lambda_1 \leq -0.4\,{\rm GeV}^2$,
which is in nice agreement with the QCD
sum rule prediction. For a further
discussion of the present status of $\lambda_1$, I refer to
\cite{NPro}.

The next step is to fix $m_b$ and $m_c$. Here one makes use of
the fact that in the framework of HQE the {\em difference} between
$m_b$ and $m_c$ is given by
\begin{equation}
m_b - m_c = m_B-m_D + \frac{\lambda_1+3\lambda_2}{2}\!\left(
\frac{1}{m_b} -\frac{1}{m_c} \right)
+ {\cal O}\!\left(\frac{1}{m^2_Q}\right)\! .\label{eq:gaehn}
\end{equation}
Thus either $m_b$ or $m_c$ remain to be fixed. Shifman et al.\
\cite{Vcb} took $m_b = (4.8\pm 0.1)\,$GeV from spectroscopy. With
$\lambda_1 = -0.5\,$GeV$^2$ and for a B lifetime $\tau_B = 1.49\,$ps,
they get $|V_{cb}| = 0.0415$. In \cite{LS,BU,LN,BN94} $m_c$ was
determined from the experimental value of $B(D\to Xe\nu)$ via Eq.\
(\ref{eq:2}) with $m_b$ replaced by $m_c$. Although this procedure has
the advantage that both $m_b$ and $m_c$ are obtained by the same
method,
the validity of the HQE for the charm quark is not beyond controversy.
The b quark mass obtained is typically larger than 5$\,$GeV and thus
considerably larger as $m_b$ from spectroscopy. On the other hand, the
b quark mass used by Shifman et al.\ seems to underestimate
 the intrinsic renormalon ambiguity mentioned in the last section.
But even using the same method, \cite{LS} and \cite{BU} obtain
different
results: $|V_{cb}| = 0.046\pm 0.008$ and $\approx 0.042$, respectively
(both numbers rescaled using $\tau_B = 1.49\,$ps). Ref.\ \cite{BN94}
takes into account the scheme--dependence of the free quark decay
contribution and using running $\overline{\rm MS}$ masses instead of
pole masses obtains $|V_{cb}| = 0.036\pm 0.005$. Since the decay rates
only differ in terms of ${\cal O}(\alpha_s^2)$, these results show
that at
present the theoretical error due to scale-- and scheme--dependence is
of paramount importance in exploiting inclusive decays of heavy mesons
and for $|V_{cb}|$ amounts to nearly 20\%.

\section{Applications II: the Semileptonic Branching Ratio of B Mesons}

The semileptonic branching ratio of B mesons is defined by
\begin{equation}\label{eq:BR}
B(B\to X e\nu) = \frac{\Gamma(B\to X e\nu)}{\Gamma_{tot}}
\end{equation}
with
\begin{equation}
\Gamma_{tot} = \sum_{\ell =
e,\,\mu,\,\tau}\!\Gamma(B\to X\ell\nu_{\ell}) + \Gamma(B\to
X_c) + \Gamma(B\to X_{c\bar c}).
\end{equation}
The explicit formulas for the decay rates can be found in
\cite{general}. The radiative corrections to ${\cal O}(\alpha_s)$
to the semileptonic decay $b\to c e\nu$ were calculated in
\cite{CM78,HP,nir}, the corrections to $b\to cud$ with a massless c
quark in
\cite{ACMP81}, with a massive c quark in \cite{tum67}; the complete
corrections to $b\to ccs$ are not known to date, but can partly be
obtained from Ref.\ \cite{HP}.

\begin{figure}[t]
\centerline{
\epsfxsize=0.45\textwidth
\epsfbox{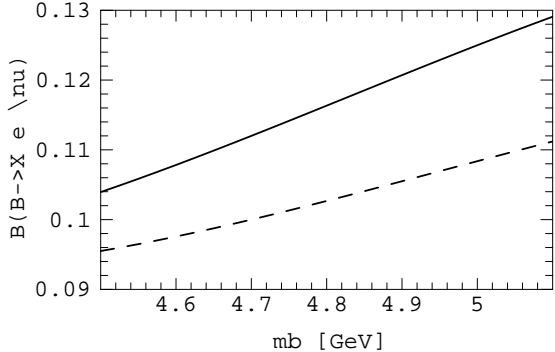}
}
\caption{$B(B\to Xe\nu)$ as function of the pole mass $m_b$ for $\mu
= m_b$. Solid line: $B$ calculated using pole masses, dashed line: $B$
calculated using running $\overline{\rm MS}$ quark masses.}
\end{figure}
\begin{table}
\begin{center}
\begin{tabular}{|c|c|c|cc|}
\hline
& Parton Model \cite{AP91} & HQE \cite{bigi} & \multicolumn{2}{c|}{HQE
\cite{tum67,tum68}} \\
$\alpha_s(m_Z)$ & pole masses & pole masses & pole masses &
$\overline{{\rm MS}}$ masses\\
\hline
0.110 & 0.132 & 0.130 & 0.121 & 0.111\\
0.117 & 0.128 & 0.126 & 0.116 & 0.103\\
0.124 & 0.124 & 0.121 & 0.111 & 0.097\\ \hline
\end{tabular}
\end{center}
\caption[]{$B(B\to Xe\nu)$ in different models depending on
$\alpha_s(m_Z)$. Input parameters: $m_b=4.8\,$GeV, $m_c = 1.3\,$GeV
(pole masses) corresponding to $\lambda_1 = -0.6\,$GeV$^2$.
Renormalization scale: $\mu = m_b$. In the phase--space
factor of $\Gamma(b\to c c s)$ $m_s=0.2\,$GeV is used. Only in
\cite{tum67,tum68} full radiative corrections to the nonleptonic B
decay modes were taken into account.}
\end{table}

The experimental branching ratios are $B(B\to Xe\nu)
= (10.43\pm 0.24)\%$ (world average) \cite{PartData} and
$B(B^0\to X e \nu) = (10.9\pm 0.7\pm 1.1)\%$, the most recent result
obtained by CLEO \cite{Cleo}.

There exists a number of theoretical analyses of $B(B\to Xe\nu)$
using different methods and obtaining
different results. In \cite{AP91}, e.g., the branching ratio was
investigated in a purely perturbative framework using the full
radiative
corrections for $b\to c e\nu$ \cite{CM78,nir}, but with $m_c=0$ in
the corrections to $b\to c ud$ and $b\to ccs$ \cite{ACMP81}. In Ref.\
\cite{bigi}, the same analysis was repeated taking into account
non--leading terms in the HQE. Finally, in \cite{tum67,tum68} also the
full radiative corrections to $b\to c ud$ were calculated and those
parts of the corrections to $b\to ccs$ available from \cite{HP}
were taken into account. Also
the effect of scheme--dependence was estimated by changing the
definition of the quark mass. For the same set of input parameters,
the results are given in Table 1. The table shows
that the introduction of nonperturbative correction terms by Bigi et
al.\ reduces $B(B\to Xe\nu)$ by 0.3\% with respect to the free quark
decay model, and that the account for the c
quark mass in the radiative corrections yields an additional --0.9\%,
which again shows the importance of {\em perturbative} corrections to
the HQE.
The scheme--dependence, however, is still tremenduous and amounts to an
uncertainty in $B(B\to Xe\nu)$ of more than 1\%. Whereas Refs.\
\cite{AP91,bigi} concluded that $B(B\to Xe\nu)>12.5\%$, Refs.\
\cite{tum67,tum68} find
\begin{equation}\label{eq:8}
B(B\to X e \nu) = (11.0\pm 1.8\pm 1.0)\%,
\end{equation}
where the first error combines uncertainties in $\alpha_s(m_Z)$,
$m_b$ (4.5$\,$GeV$\,<m_b<\,$5.1$\,$GeV), the hadronic corrections, the
renormalization scale and the uncertainty in the radiative corrections
to $b\to ccs$.
The second error is a ``guestimate'' of the theoretical error due to
scheme--dependence. The combined effect of complete radiative
corrections, new results on $\alpha_s(m_Z)$ \cite{bethke} and the
consideration of different definitions of the quark mass
thus lowers the theoretical branching ratio, which now agrees with the
experimental one within the errors and seriously restricts any possible
``new physics'' in nonleptonic B decays. A more detailed analysis
is in preparation \cite{tum68}.

\end{document}